\documentclass[a4paper,11pt]{article}
\pdfoutput=1
\usepackage{jinstpub} % for details on the use of the package, please
\title{\boldmath Study of silicon sensors for precise timing measurement}

\author[a,1]{Y. Deguchi,\note{Corresponding author.}}
\author[a]{K. Kawagoe,}
\author[a]{R. Mori,}
\author[c]{E. Mestre,}
\author[a]{T. Suehara,}
\author[b]{T. Yoshioka,}

\affiliation[a]{Kyushu University,\\744 Motooka Nishi-ku Fukuoka, Japan}
\affiliation[b]{Research Center for Advanced Particle Physics, Kyushu University,\\744 Motooka Nishi-ku Fukuoka, Japan}
\affiliation[c]{Universite\'{e} PARIS-SACLAY,\\91190 Saint-Aubin, France}
\emailAdd{deguchi@epp.phys.kyushu-u.ac.jp}

\abstract{Silicon sensors with high time resolution can help particle identification in the International Linear Collider (ILC). We are studying Low Gain Avalanche Diodes (LGADs) as a high timing resolution sensor. As a step to develop LGADs, we are now focusing to characterize Avalanche Photo Diode (APD)s, because the APDs has the same multiplication structure as LGADs. We studied the characteristics of APDs with particles from radioisotopes.}


\proceeding {Calorimetry for the High Energy Frontier, CHEF 2019\\
  25-29 November\\
  Fukuoka, Japan}

\begin{document}
\maketitle
\flushbottom

\section{Introduction} \label{sec:intro}
The International Linear Collider (ILC)\cite{design} is an electron-positron linear collider, which is planned to be constructed in Japan. The main purposes of the ILC are discoveries of new particles and precise measurement of the Higgs boson and the top quark. The center of mass energy will be 250 GeV with 20 km tunnel length. In the future, the center of mass energy can be upgraded to 1 TeV with 50 km tunnel length. \par
In the International Large Detector (ILD), one of two detector concepts for the ILC, particle identification can be achieved with dE/dx and momentum measurement with the Time Projection Chamber (TPC), but it has insensitive energy range which gives similar dE/dx with multiple kinds of particles. With a precise Time-of-Flight (ToF) measurement combined with the dE/dx measurement, performance on the particle identification can be improved. In order to identify pions, kaons and protons up to 5 GeV/$c$, the time resolution is required to be less than 50 psec. We are studying Low Gain Avalanche Diodes (LGADs) to achieve the resolution. \par
In this paper, as a step to develop LGADs, we are now focusing to characterize Avalanche Photo Diode (APD)s, which are usually used to measure optical photons, with charged particles. Since the APDs has the same multiplication structure as LGADs, this should help determining the structure of the LGADs. We studied the characteristics of APDs with particles from radioisotopes.

\section{Particle identification} \label{sec:partcle_id}
\begin{figure}[h]
 \centering 
 \includegraphics[width=8cm]{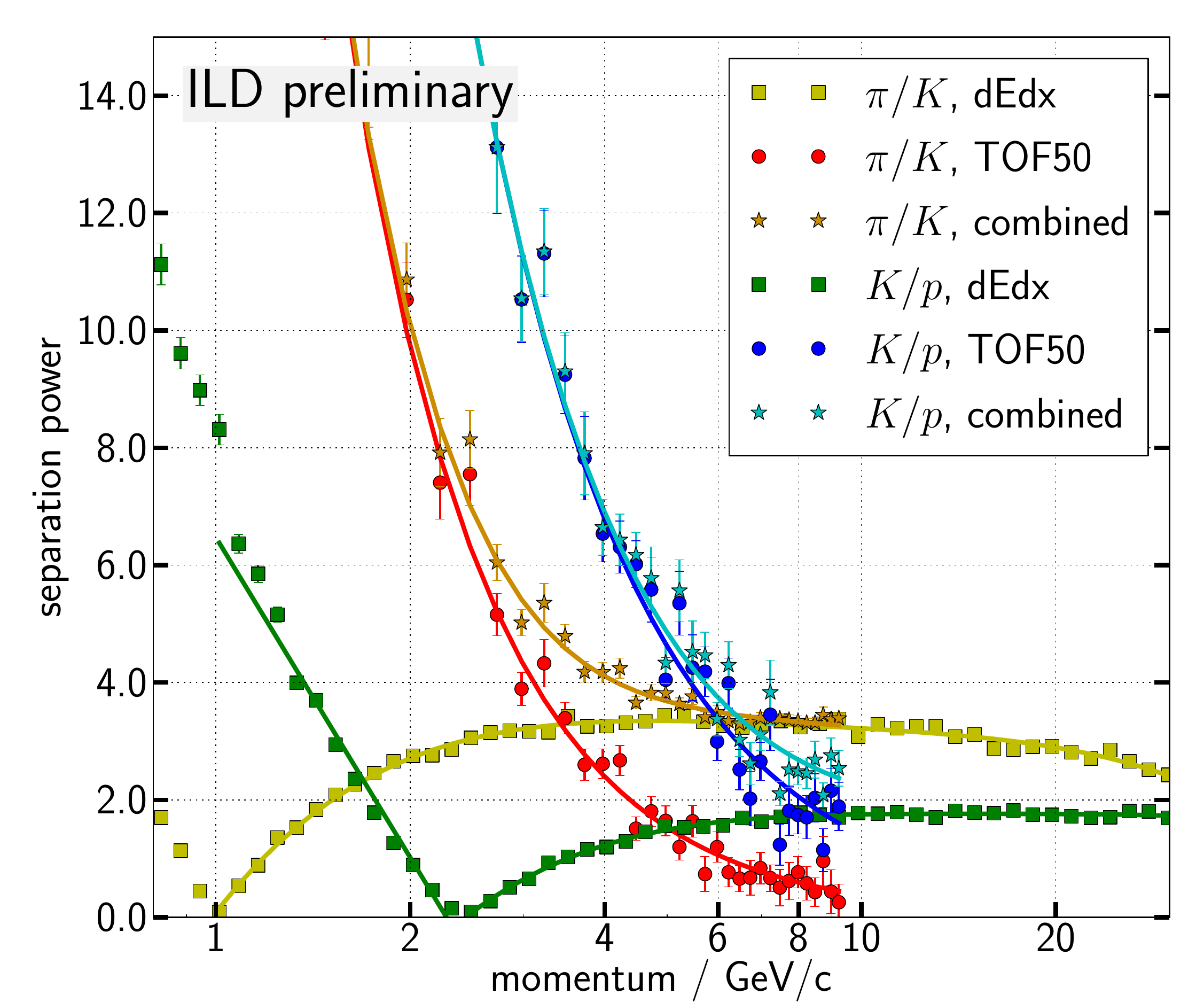}
 \caption{Separation power of particles for the TPC of ILD. In the identification of $\pi$ and $K$, these line shows the separation with only TPC (yellow), with only ToF (red), and with combined TPC and ToF (orange). In the identification of $K$ and $p$, these line shows the separation with only TPC (green), with only ToF (blue), and with TPC and ToF (light blue) combined.}
 \label{fig:dedx_tof}
\end{figure}

%In the ILD, particle identification can be achieved with dE/dx and momentum measurement with TPC, but there is a insensitive energy range with only TPC. 
Figure~\ref{fig:dedx_tof} shows the separation power of particles for the TPC of ILD. The yellow and green line show the separation power between $\pi/K$ and between $K/p$ with only TPC. There is a point where the separation power is 0 on each line. Then we suggest the particle identification with ToF. Particles have differences of flight time depending on their mass. The orange line and light blue line show the separation power between $\pi/K$ and between $K/p$ with dE/dx and ToF combined. To achieve the separation power up to 5 GeV/$c$, the time resolution down to 50 ps is needed.

\section{Silicon Sensors} \label{sec:sensor}
\subsection{LGAD} \label{subsec:lgad}
LGADs are silicon sensors with the internal amplification mechanism, which already have been proved to realize the time resolution down to 30 psec \cite{lgad}. However, the normal reach-through LGADs have an issue that the amplification factor heavily depends on the position of the hit, because the amplification region is not uniformly formed due to the surface structure. To overcome this, inverse-type LGADs have been proposed, which has amplification region at the bottom, in contrast to the reach-through type with amplification occurring just below the surface. 
 
\subsection{APD} \label{subsec:apd}
APDs are silicon sensors known as photon detectors, which have the same structure as LGADs.  Table~\ref{tab:apd_used} shows the list of APD owned by Kyushu University, made by Hamamatsu. The pkg-10 and pkg-20 are LGAD prototype provided for Hamamatsu. In this paper, we used these APDs to study these characteristics.

\begin{table}[h]
\centering
\caption{The list of APD owned by Kyushu Univ.}
\label{tab:apd_used}
 \begin{tabular}{c|crr} \hline
   APD serial No. & type & \multicolumn{1}{c}{$V_{{\rm br}}$} & \multicolumn{1}{c}{active area} \\ \hline \hline
   S12023-10A & reach-through & 139 V& $\phi\ 1\ $mm \\  
   S8664-10K & inverse & 417 V & $\phi\ 1\ $mm \\ 
   pkg-10 & reach-through & about 250 V & $\phi\ 1\ $mm \\ 
   pkg-20 & reach-through & about 120 V & $\phi\ 1\ $mm \\ 
   S3884 & reach-through & 159 V & $\phi\ 1.5\ $mm \\ 
   S2384 & reach-through & 189 V & $\phi\ 3\ $mm \\ 
   S8664-20K & inverse & 425 V & $\phi\ 2\ $mm \\ 
   S8664-55 & inverse & 433 V & $5\times5\ \rm{mm^2}$ \\ \hline
 \end{tabular}
\end{table}

\section{Measurement using radioisotopes} \label{sec:measurement}
\subsection{Setup} \label{subsec:setup}
\begin{figure}[h]
 \centering 
 \includegraphics[width=9cm]{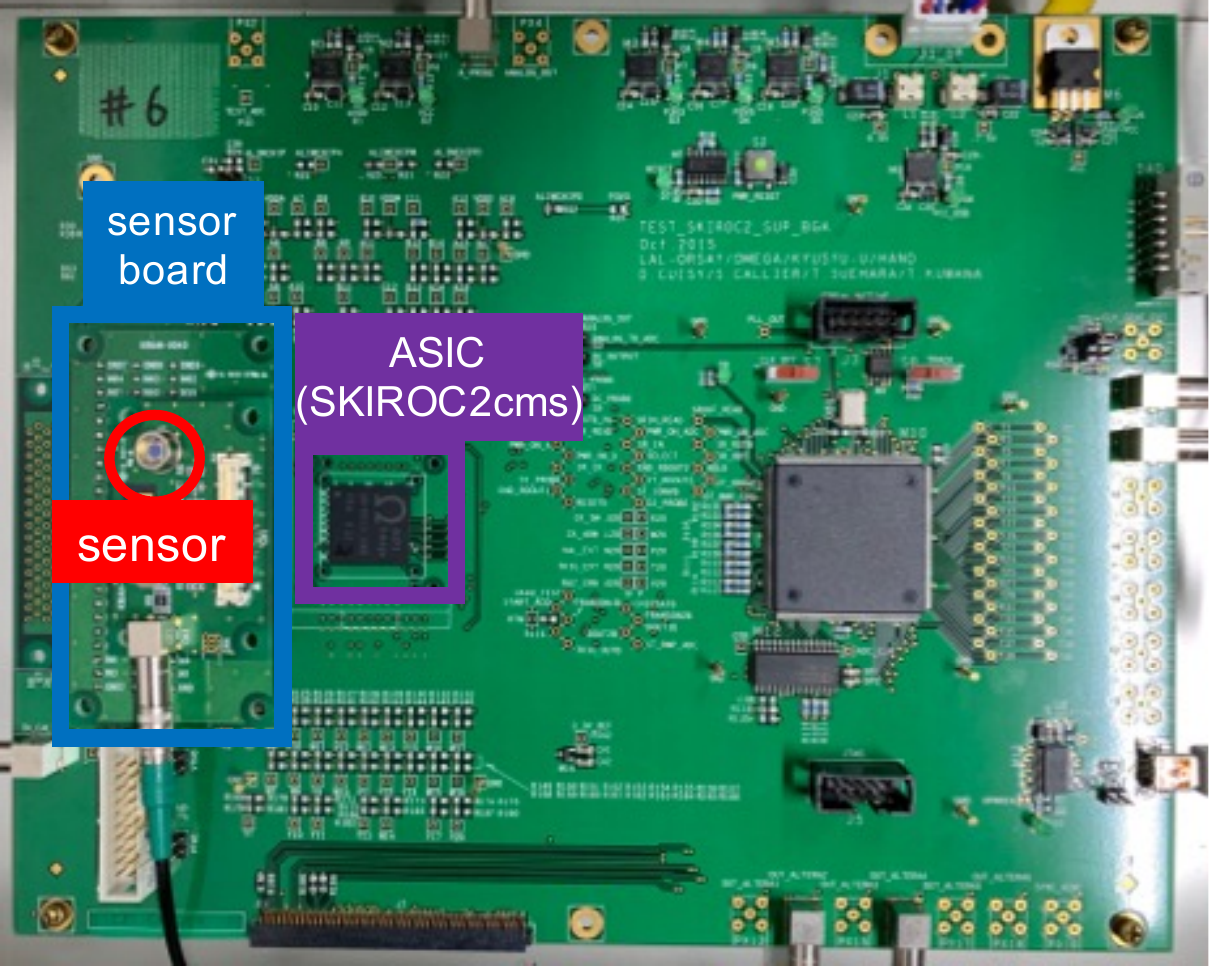}
 \caption{Testboard (with APD connected and which soldered SKIROC2cms)}
 \label{fig:testboard}
\end{figure}

We used the Testboard and SKIROCcms to read out signals from APDs. Testboard is a board to evaluate the characteristics of ASICs. It can be connected to APDs using the sensor boards as shown in Fig~\ref{fig:testboard} and acquire data from APDs. The SKIROC2cms is an ASIC which can perform charge and time measurements. In these measurements, we put these equipments into an aluminum box for noise shielding. \par
We used a ${}^{90}\rm{Sr}$ beta source and a ${}^{133}\rm{Ba}$ gamma source. The beta source is used to observe the behavior when a MIP enters the APDs. The gamma source is used to measure the gain of APDs. 

\subsection{$\beta$ source measurement} \label{sebsec:beta}
\begin{figure}[h]
 \centering 
 \includegraphics[width=9cm]{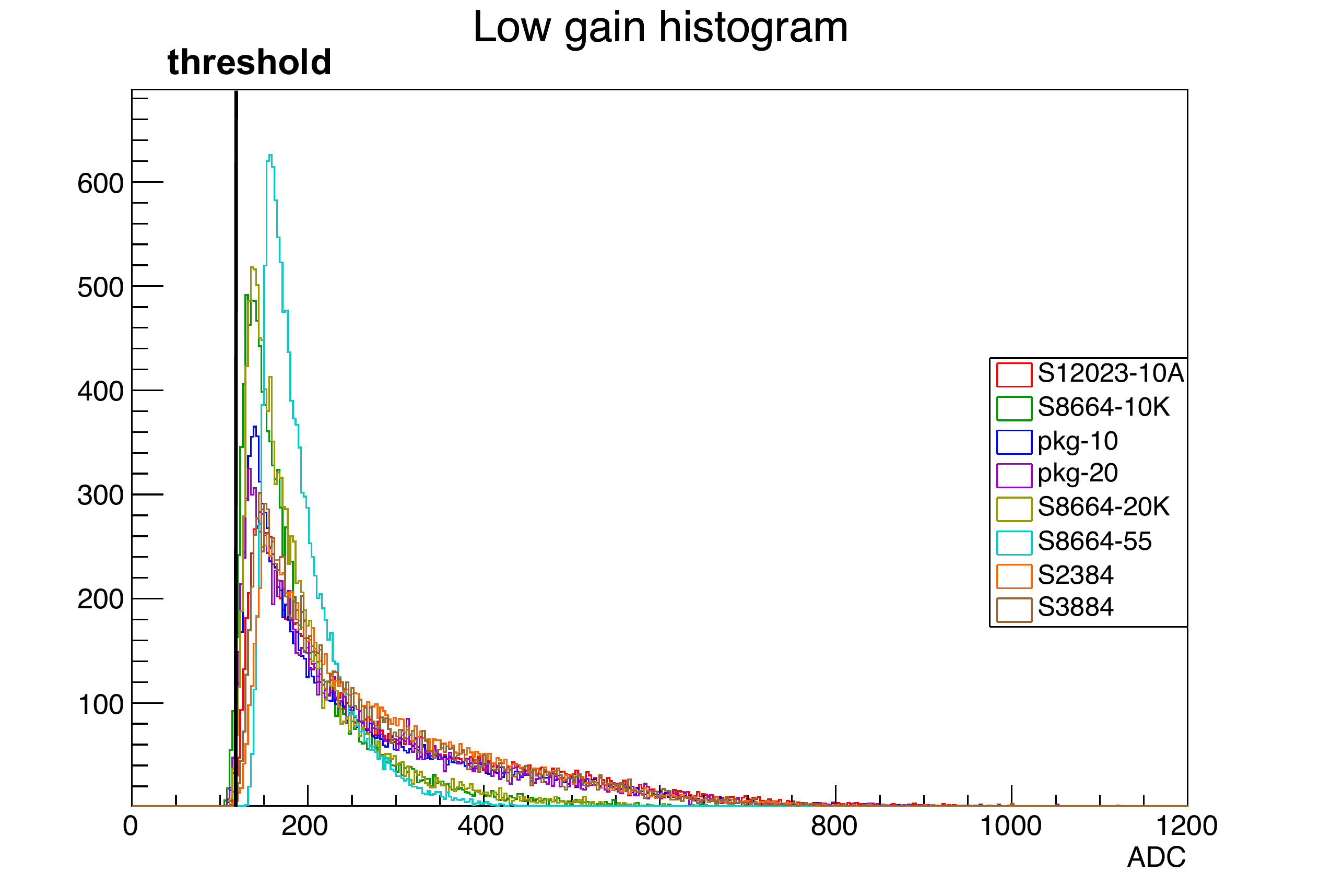}
 \caption{ADC histograms using ${}^{90}\rm{Sr}$}
 \label{fig:beta_hist}
\end{figure}

Figure~\ref{fig:beta_hist} shows ADC histograms of APDs using beta source. These histograms showed that the signal height of reach-through type APDs is slightly larger than inverse type's one, and S8664-55, which has the most largest active area, is larger than other APDs. 

\begin{figure}[h]
 \centering 
 \includegraphics[width=9cm]{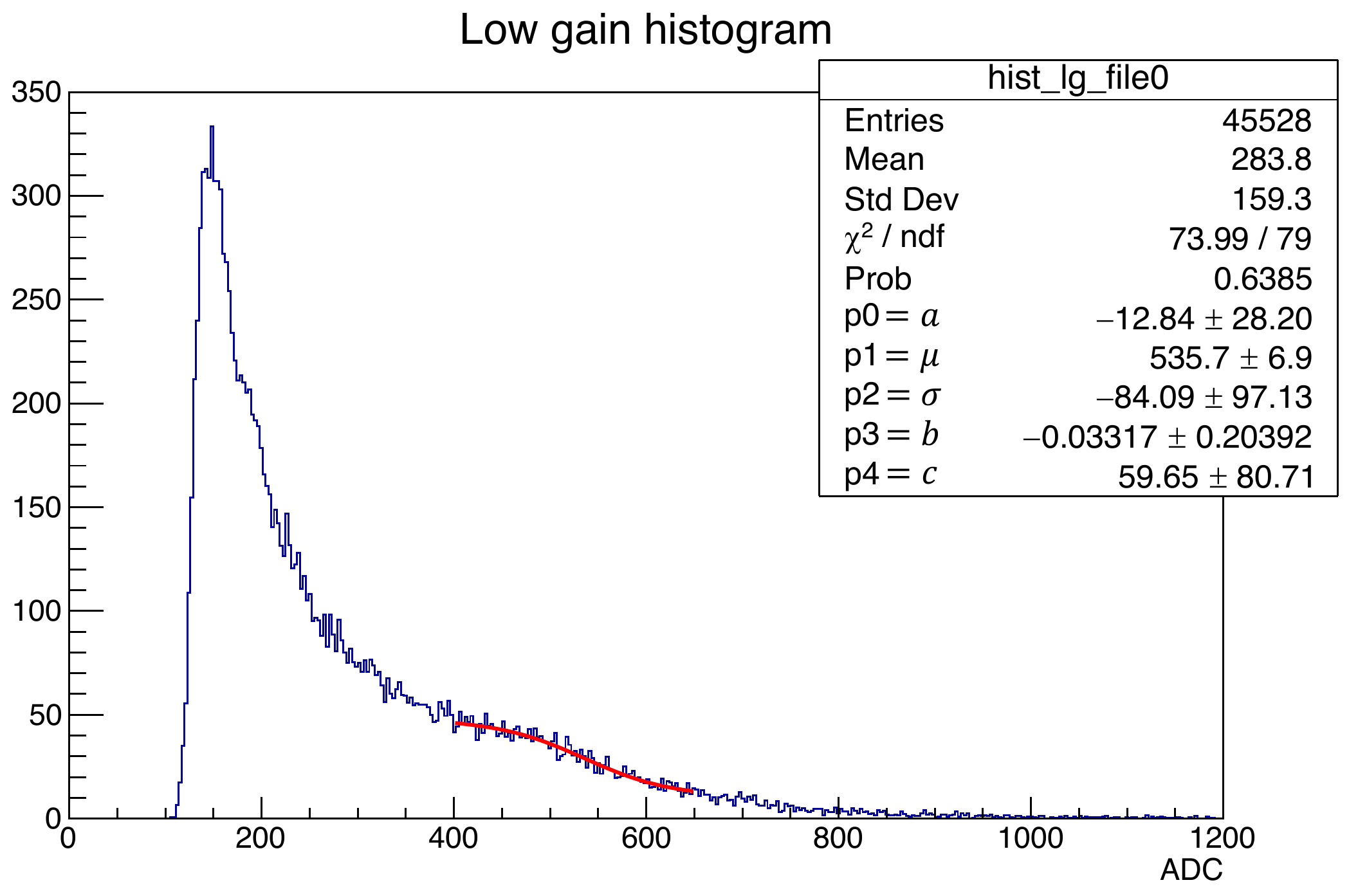}
 \caption{ADC histogram of S12023-10A using ${}^{90}\rm{Sr}$. }
 \label{fig:beta_hist_one}
\end{figure}

In the histogram of S12023-10, there is a small shoulder (Fig.~\ref{fig:beta_hist_one}, around 500 ADC counts). The ADC distribution should be landau distribution when the MIPs enter the APD, but if the APD has gain variation, the ADC distribution will be integration landau distributions. Assuming this ADC distribution is integration of landau distributions, the shoulder should be at the landau distribution of maximum gain. We fitted the peak using a function combing an error function and a linear function, as 
\begin{equation} \label{eq:error}
f(x)=\frac{2a}{\sqrt{\pi}}\int_{\frac{x-\mu}{\sigma}}^{\infty}e^{-t^2}dt+bx+c
\end{equation}
The $\mu$ and $\sigma$ are the position and width of error function. We used the position of this shoulder to calculate active thickness of S12023-10A.

\subsection{$\gamma$ source measurement} \label{subsec:gamma}
In order to calculate the active thickness, we have to measure the gain of APD. In this paper, we measured the gain of S12023-10A using gamma source. \par
\begin{figure}[h]
 \begin{tabular}{cc}
  \begin{minipage}{0.5\hsize}
   \centering 
   \includegraphics[width=7.5cm]{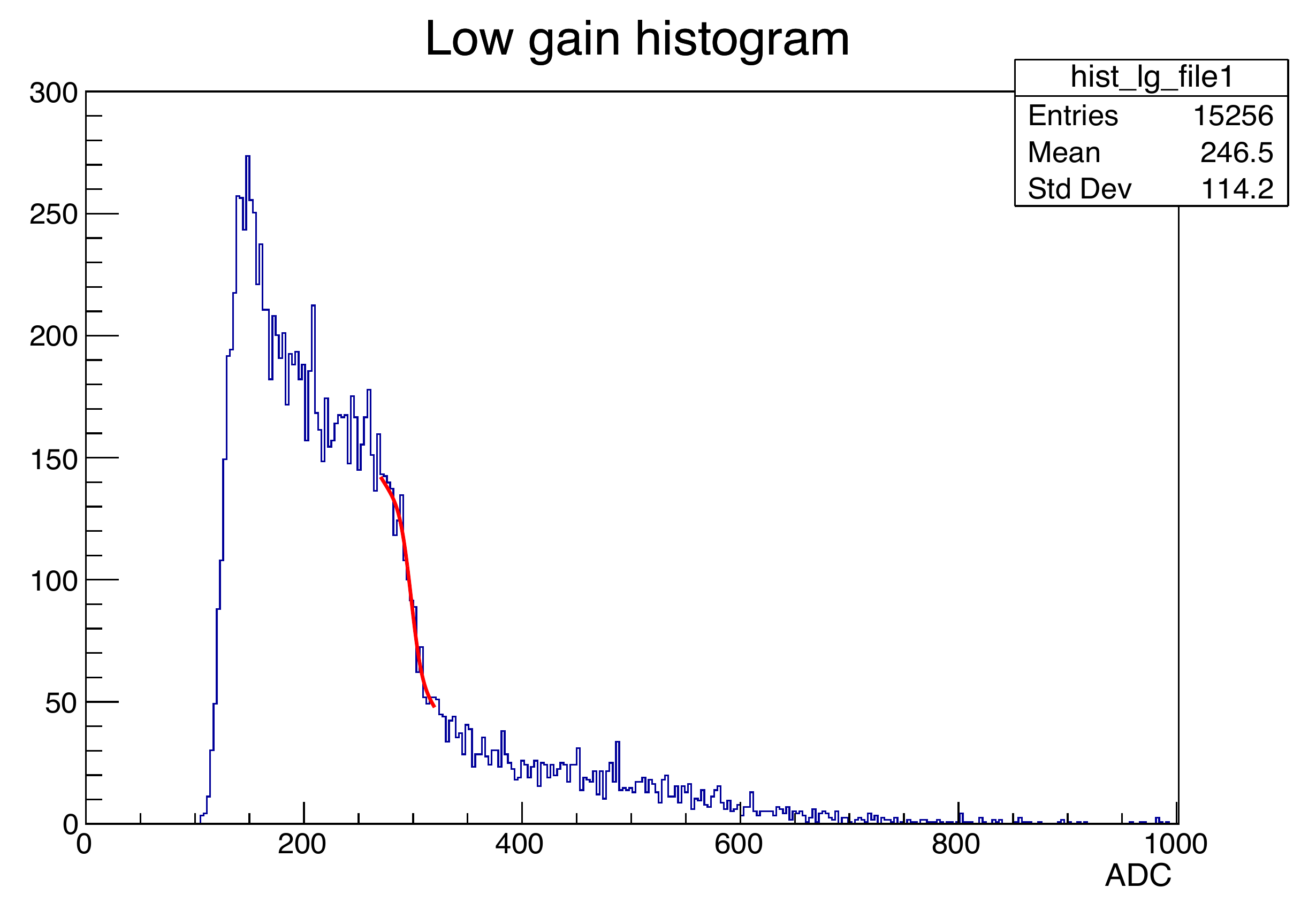}
  \end{minipage}
  \begin{minipage}{0.5\hsize}
   \centering
   \includegraphics[width=7.5cm]{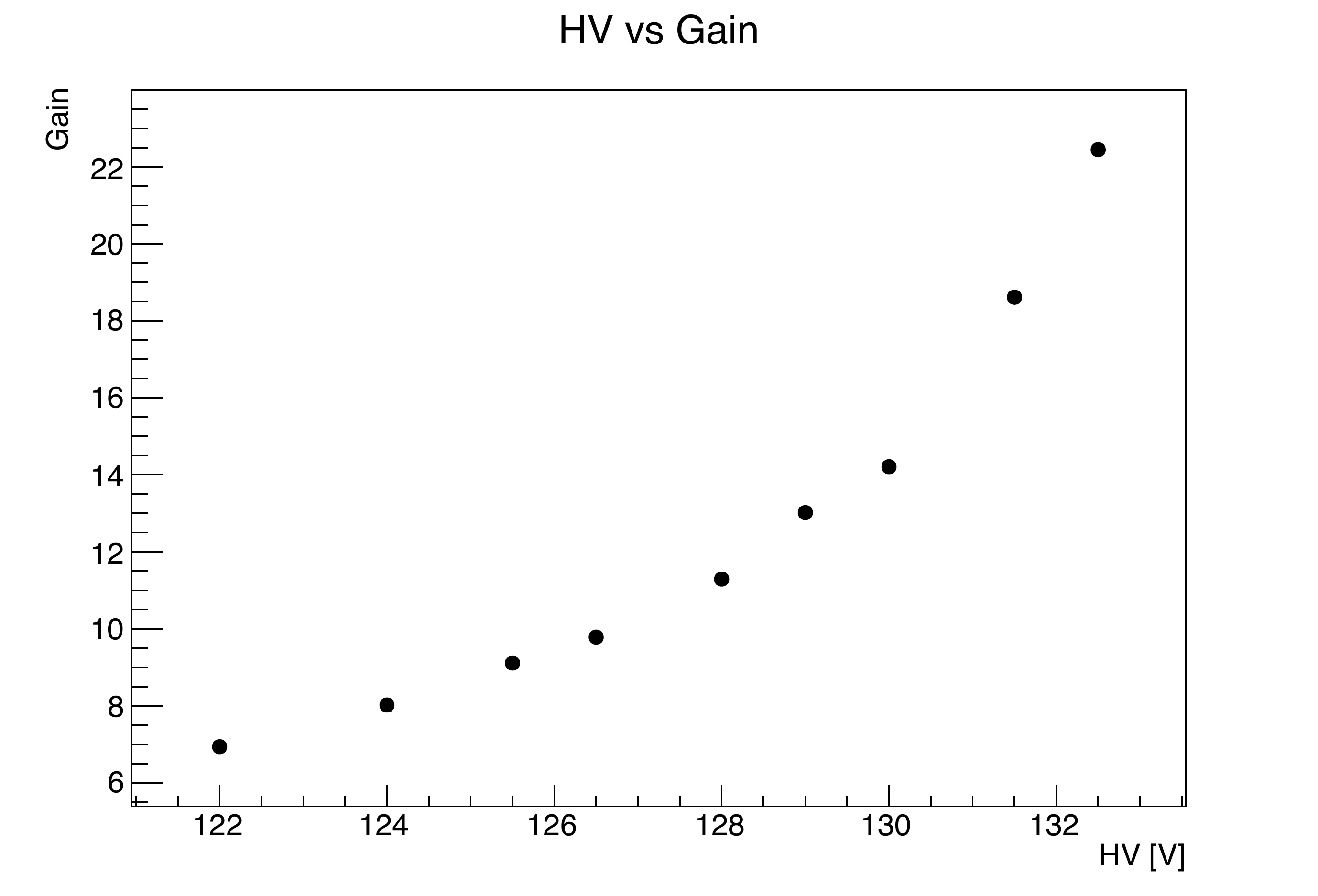}
  \end{minipage}
 \end{tabular}
 \caption{The result of gain measurement using ${}^{133}\rm{Ba}$}
 \label{fig:gamma_plot}
\end{figure}

 
Figure~\ref{fig:gamma_plot} (left) shows the ADC distribution of S12023-10A, with the bias voltage of 129 V. The edge of ADC distribution, at about 300 ADC counts, is due to the Compton edge, because the spectrum of ${}^{133}\rm{Ba}$ has a peak of photoelectric effect at 356 keV, and the Compton edge is at 207 keV. We fitted this edge using the same function as beta source measurement \eqref{eq:error}, and calculated the gain using the mean of this function.  The voltage dependence of the position of the shoulder is shown in Figure~\ref{fig:gamma_plot} (right).

\subsection{Active thickness} \label{subsec:active}
We calculated the active thickness of S12023-10A using the results of beta and gamma sources measurements. A MIP makes 76 electron and hole pairs per 1 ${\rm \mu}$m in a silicon sensor, so the formula for calculating active thickness can be described by 
\begin{equation} \label{eq:active}
{\rm Active\ thickness}=\frac{\mu}{a}\cdot\frac{1}{{\rm Gain}}\cdot\frac{1}{76\cdot e}
\end{equation}
with the $\mu$ is the mean of error function when measured using beta source and $e=1.602\times10^{16}$ C.  The $a$ is gradient of the correlation between ADC and charge in SKIROC2cms. Finally, the active thickness of S12023-10A is $\sim 1000 {\rm \mu m}$.  \par
However, it is said that the active thickness of APDs is a few tens of ${\rm \mu}$m and the gain of APD at 129 V is about 500 (according to the datasheets).  Assuming the gain value is 500, the active thickness is $\sim 30{\rm \mu m}$ calculated by \eqref{eq:active}.

\section{Summary} \label{sec:summary}
We are studying LGAD to identify the charged particles in ILD. As a step to develop LGADs, we are now focusing to characterize APDs. We measured the characteristics when MIPs enter the APDs using beta source, and estimated the active thickness of S12023-10A. We expect to measure the active thickness of inverse type APD, and provide LGAD prototype for ILC.

%\appendix
%\section{Some title}
%Please always give a title also for appendices.

\acknowledgments
This wark is supported by Omega group. This work was supported by JSPS KAKENHI Grant Number JP17H05407.

%\paragraph{Note added.} This is also a good position for notes added
%after the paper has been written.


\end{document}